\documentclass[conference]{IEEEtran}
\usepackage{cite}
\usepackage{amsmath,amssymb,amsfonts}
\usepackage{algorithmic}
\usepackage{graphicx}
\usepackage{textcomp}
\usepackage{xcolor}
\usepackage[utf8]{inputenc} 
\usepackage{booktabs}       
\usepackage{amsfonts}       
\usepackage{nicefrac}       
\usepackage{microtype}      
\usepackage{amsmath}
\usepackage{graphicx}
\graphicspath{ {images/} }
\usepackage{hyperref}       
\usepackage{subfig}
\usepackage{float}
\usepackage{textcmds}
\usepackage{blindtext}
\usepackage{caption}

\author{\textbf{Shouvik Roy}\\ \textit{Stony Brook University}\\ \texttt{shroy@cs.stonybrook.edu} \\
\and 
\textbf{Usama Mehmood}\\ \textit{Stony Brook University}\\ \texttt{umehmood@cs.stonybrook.edu} \\
  \and 
\textbf{Radu Grosu}\\ \textit{Vienna University of Technology}\\ \texttt{radu.grosu@tuwien.ac.at} \and \\
\textbf{Scott A. Smolka}\\ \textit{Stony Brook University}\\ \texttt{sas@cs.stonybrook.edu}  \and \\
\textbf{Scott D. Stoller}\\ \textit{Stony Brook University}\\ \texttt{stoller@cs.stonybrook.edu}  \and \\
\textbf{Ashish Tiwari}\\ \textit{Microsoft Research}\\ \texttt{ashish.tiwari@microsoft.com}}

\begin{document}

\title{Neural Flocking: MPC-based Supervised Learning\\ of Flocking Controllers}

\maketitle

\begin{abstract}
We show how a distributed flocking controller can be synthesized using deep learning from a centralized controller which generates the trajectories of the flock.  Our approach is based on \emph{supervised learning}, with the centralized controller providing the training data to the learning agent, i.e., the synthesized distributed controller.  We use Model Predictive Control (MPC) for the centralized controller, an approach that has been successfully demonstrated on flocking problems.  MPC-based flocking controllers are high-performing but also computationally expensive.  By learning a symmetric distributed neural flocking controller from a centralized MPC-based flocking controller, we achieve the best of both worlds: the neural controllers have high performance (on par with the MPC controllers) and high efficiency.  Our experimental results demonstrate the sophisticated nature of the distributed controllers we learn.  In particular, the neural controllers are capable of achieving myriad flocking-oriented control objectives, including flocking formation, collision avoidance, obstacle avoidance, predator avoidance, and target seeking.  Moreover, they generalize the behavior seen in the training data in order to achieve these objectives in a significantly broader range of scenarios.
\end{abstract}


\section{Introduction}

With the introduction of Reynolds rule-based model~\cite{REYNOLDS1987,Reynolds99}, it is now possible to understand the flocking problem as one of distributed control.
Specifically, in this model, at each time-step, each agent executes a control law given in terms of the weighted sum of three competing forces to determine its next acceleration.  Each of these forces has its own rule: \emph{separation} (keep a safe distance away from your neighbors), \emph{cohesion} (move towards the centroid of your neighbors), and \emph{alignment} (steer toward the average heading of your neighbors). Reynolds controller is \emph{distributed}, i.e., it is executed separately by each agent,  using information about only itself and nearby agents, and without communication.  Furthermore, it is \emph{symmetric}; i.e., every agent runs the same controller (same code). It was subsequently shown that a simpler, more declarative approach to the flocking problem is possible \cite{Mehmood2018}.  In this setting, flocking is achieved when the agents combine to minimize a flock-wide cost function.  Centralized and distributed solutions for achieving this form of ``declarative flocking'' were presented, both of which were formulated in terms of Model-Predictive Control (MPC) \cite{CAMACHO2007}. The problem with MPC is that computing the next control action can be computationally expensive, as MPC attempts to find an action that minimizes the cost function over a given prediction horizon. This renders MPC unsuitable for real-time applications with short control periods, for which flocking is a prime example.  Another potential problem with MPC-based approaches to flocking is its performance (at achieving the desired flight formations), which may suffer in a fully distributed setting.

In this paper, we present \emph{Neural Flocking (NF)}, a new approach to the flocking problem that uses Supervised Learning to learn a symmetric and fully distributed flocking controller from a centralized MPC-based controller.  By doing so, we achieve the best of both worlds: high performance (on par with the MPC controllers) in terms of meeting flocking flight-formation objectives, and high efficiency leading to real-time flight controllers. Moreover, our NF controllers can easily be parallelized on specialized hardware such as GPUs and TPUs.

\begin{figure*}[t!]
    \centering
    \includegraphics[width=16cm]{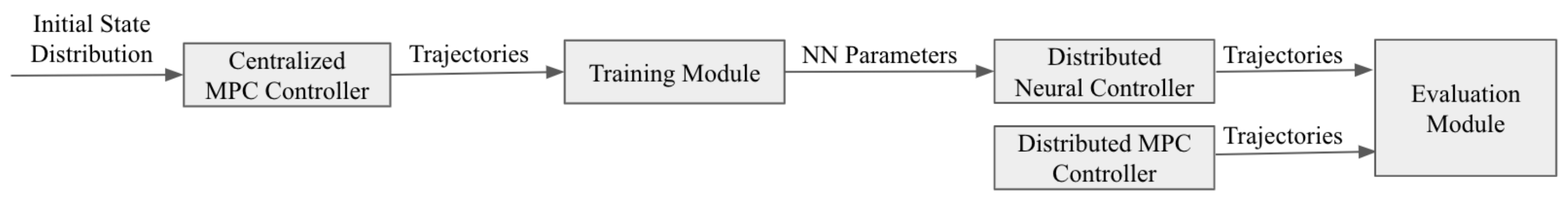}
    \caption{Neural Flocking Architecture}
    \label{fig:architecture}
\end{figure*}

Figure~\ref{fig:architecture} gives an overview of the NF approach.  A high-performing centralized MPC controller provides the labeled training data to the learning agent: a symmetric and distributed neural controller in the form of a DNN or LSTM. The training data consists of trajectories of state-action pairs, where a state contains the information known to an agent at a time step (e.g., its own position and velocity, and the position and velocity of its neighbors), 
and the action (the label) is the acceleration assigned to that agent at that time step by the centralized MPC controller.

We formulate and evaluate NF in a number of essential flocking scenarios: basic flocking as in~\cite{REYNOLDS1987,Mehmood2018}, and more advanced flocking scenarios with additional objectives, including inter-agent collision avoidance, obstacle avoidance, predator avoidance, and target seeking by the flock. We conduct an extensive performance evaluation of NF.  Our experimental results, which include videos, demonstrate the sophisticated nature of NF controllers.  In particular, they are capable of achieving all of the control objectives listed above. Moreover, they generalize the behavior seen in the training data in order to achieve these objectives in a significantly broader range of scenarios.
\section{Background}
\label{sec:bg}

We consider a set of $n$ dynamic agents $\mathcal{A} = \{1,\ldots,n\}$ that move according to the following discrete-time equations of motion:
\begin{equation} \label{motioneq}
\begin{split}
    p_i(k+1) &= p_i(k) + dt \cdot v_i(k),\quad |v_i(k)| < \bar{v} \\
    v_i(k+1) &= v_i(k) + dt \cdot a_i(k),\quad |a_i(k)| < \bar{a}   
\end{split}
\end{equation}
where $p_i(k) \in \mathbb{R}^2$, $v_i(k) \in \mathbb{R}^2$, $a_i(k) \in \mathbb{R}^2$ are the position, velocity and acceleration of agent $i \in \mathcal{A}$ respectively at time step $k$, and $dt \in \mathbb{R}^+$ is the time step. The magnitudes of velocities and accelerations are bounded by $\bar{v}$ and $\bar{a}$, respectively.  We define the state of agent $i$ as the set $s_i = \{p_i, v_i\} $. 
Acceleration $a_i(k)$ is the control input for agent $i$ at time step $k$. The acceleration is updated after every $\eta$ time steps, where $\eta$ is the duration of the control step relative to the time step.
The flock \emph{configuration} at time step $k$ is thus given by the following vectors (in boldface):
\begin{align} \label{flockconfig}
    \textbf{p}(k) &= [p_1^T(k) \cdot \cdot \cdot p_n^T(k)]^T \\
    \textbf{v}(k) &= [v_1^T(k) \cdot \cdot \cdot v_n^T(k)]^T \\
    \textbf{a}(k) &= [a_1^T(k) \cdot \cdot \cdot a_n^T(k)]^T 
\end{align}

The configuration vectors are referred to without the time indexing as $\textbf{p}$, $\textbf{v}$, and $\textbf{a}$. The \emph{neighborhood} of agent $i$ at time step $k$, denoted by $\mathcal{N}_i(k) \subseteq \mathcal{A}$, contains its $N$-nearest neighbors, i.e., the $N$ other agents closest to it.  We use this definition for simplicity, and expect that a radius-based definition of neighborhood would lead to similar results for our distributed flocking algorithms. 

\subsection{Model-Predictive Control}
\label{sec:mpc}

Model-Predictive control (MPC) \cite{CAMACHO2007} is a well-known control technique that has recently been applied to the flocking problem~\cite{Mehmood2018, zhang2015model, zhan2013flocking}.  At each control step, an optimization problem is solved to find the optimal sequence of control actions (agent accelerations in our case) that minimizes a given cost function with respect to a predictive model of the system.  The first control action of the optimal control sequence is then applied to the system; the rest is discarded. In the computation of the cost function, the predictive model is evaluated for a finite prediction horizon of $T$ control steps.

MPC-based flocking models can be categorized as \textit{centralized} or \textit{distributed}. A \textit{centralized} model assumes that complete information about the flock is available to the central controller, which uses the states of all agents to compute the optimal accelerations for each agent.
The following optimization problem is solved by a centralized MPC at each control step $k$: 
\begin{equation}
\label{eq:central_mpc}
\underset{\mathbf{a}(k \mid k), \ldots, \mathbf{a}(k+T-1 \mid k) \,<\, \bar{a}}{\min} \quad J(k) + \lambda \cdot \sum_{t=0}^{T-1} \| \mathbf{a}(k + t\mid k) \|^2 
\end{equation}
The first term $J(k)$ is the centralized model-specific cost, evaluated for $T$ control steps (this embodies the predictive aspect of MPC), starting at time step $k$. It encodes the control objective. The second term, scaled by a weight $\lambda > 0$, penalizes large control inputs: $\mathbf{a}(k+t \mid k)$ are the predictions made at time step $k$ for the accelerations at time step $k + t$. 

In \emph{distributed MPC}, each agent computes its acceleration based only on its state and its local knowledge, e.g., information about its neighbors: 

\begin{equation}
\label{eq:distr_mpc}
\underset{a_i(k \mid k), \ldots, a_i(k+T-1 \mid k) \,<\, \bar{a}}{\min} \quad J_i(k) + \lambda \cdot \sum_{t=0}^{T-1} \| {a_i(k + t\mid k)}\|^2 
\end{equation}
$J_i(k)$ is the distributed model-specific cost function for agent $i$, analogous to $J(k)$. In distributed MPC, due to limited information, an agent cannot calculate the exact future behaviour of its neighbors. Hence, the predictive aspect of $J_i(k)$ must rely on some assumption about that behavior during the prediction horizon.  Our distributed cost functions are based on the assumption that the neighbors have zero accelerations during the prediction horizon.  While this simple design is clearly not completely accurate, our experiments show that it still achieves good results.

\subsection{Declarative Flocking}
\label{sec:dflock}

Declarative flocking (DF) \cite{Mehmood2018} is a high-level approach to designing flocking algorithms, by defining a suitable cost function for MPC, instead of defining the algorithms operationally using rules, as in Reynolds model. For basic flocking, the cost function contains two terms: (1)~A \emph{cohesion} term based on the squared distance among all pairs of agents in the flock; and (2)~a \emph{separation} term based on the inverse of the squared distance among the agents. The flock evolves toward a configuration in which these two opposing forces are balanced.   For centralized DF, i.e., centralized MPC (CMPC), the cohesion term considers all pairs of agents, and the separation term considers only neighbors. 

\begin{multline} 
\label{eq_cmpc}
J^C \left(\mathbf{p}\right) = \frac{2}{|\mathcal{A}|\cdot(|\mathcal{A}| - 1)} \cdot \sum_{i\in \mathcal{A}}{\sum_{j \in \mathcal{A}, i < j}{\|p_{ij}\|^2}} \\ +  \omega \cdot \sum_{(i,j) \in \mathcal{E}(\mathbf{x})} \frac{1}{\|p_{ij}\|^2}
\end{multline}
where $\omega$ is the weight of the separation term and controls the density of the flock, and $\mathcal{E}(\mathbf{p})$ $=$ $\left\lbrace(i,j) \in \mathcal{A}\times\mathcal{A} \mid \| p_i - p_j \| < r, i\neq j\right\rbrace$ is the set of pairs of agents separated by distance less than $r$, where $r$ defines the distance-based neighborhood.
The control law for CMPC is given by Eq.~(\ref{eq:central_mpc}), with $J(k) = \sum_{t=1}^T{J^{\rm C}\left(\mathbf{p}(k+t \mid k)\right)}$. 

The basic flocking cost function for distributed DF is similar to that for CMPC, except that the cost function $J^D_i$ for agent $i$ is computed over its set of neighbors $\mathcal{N}_i$:
\begin{equation}
\label{eq_dmpc}
J^{\rm D}_i\left(\mathbf{p}\right) = \frac{1}{|\mathcal{N}_i(k)|} \cdot \sum_{j \in \mathcal{N}_i(k)}{\|p_{ij}\|^2} +  \omega \cdot \sum_{j \in \mathcal{N}_i(k)} \frac{1}{\|p_{ij}\|^2}
\end{equation}
The control law for agent $i$ is Eq.~(\ref{eq:distr_mpc}), with $J_i(k)\,{=}\,\sum_{t=1}^T{J^{\rm D}_i\left(\mathbf{p}(k+t \mid k)\right)}$.

\section{Neural Flocking}
\label{sec:nflock}
We learn a \emph{distributed neural controller} (DNC) from trajectories obtained from a CMPC. In addition to learning basic flocking behavior, we learn additional flocking-related behaviors, namely, collision avoidance, obstacle avoidance, target seeking, and predator avoidance. 
We also show how the learned behavior generalizes over a larger number of agents to achieve a successful collision-free flocking.

We use \emph{supervised learning} to train our DNC. Supervised learning learns a function that maps an input to an output based on an example sequence of input-output pairs. For our task, the trajectory data obtained from CMPC contains both the training inputs and training labels: the state of the agent is the input, and the agent's acceleration at the same time step is the label, i.e., the output.

\subsection{Required Extensions to Declarative Flocking}
\label{sec:df_extensions}
\begin{figure}[t]
\centering
\subfloat[Basic flocking]{\includegraphics[width=.24\textwidth]{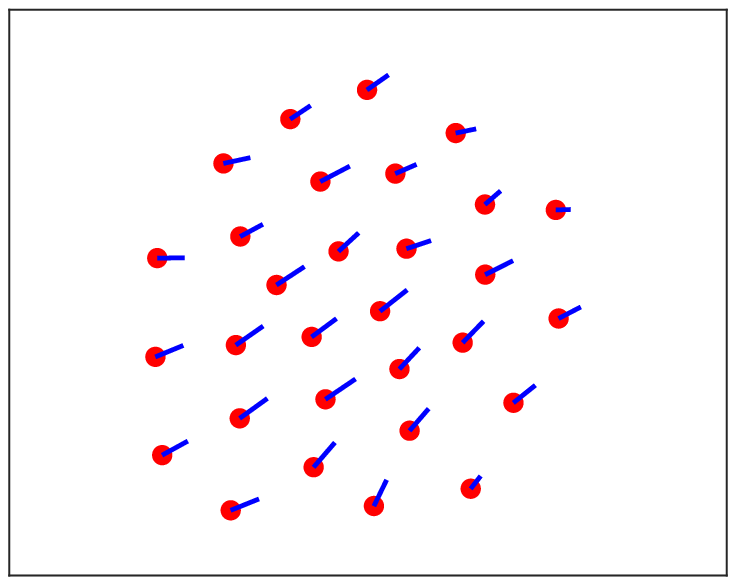}}\hfill
\subfloat[Obstacle avoidance]{\includegraphics[width=.24\textwidth]{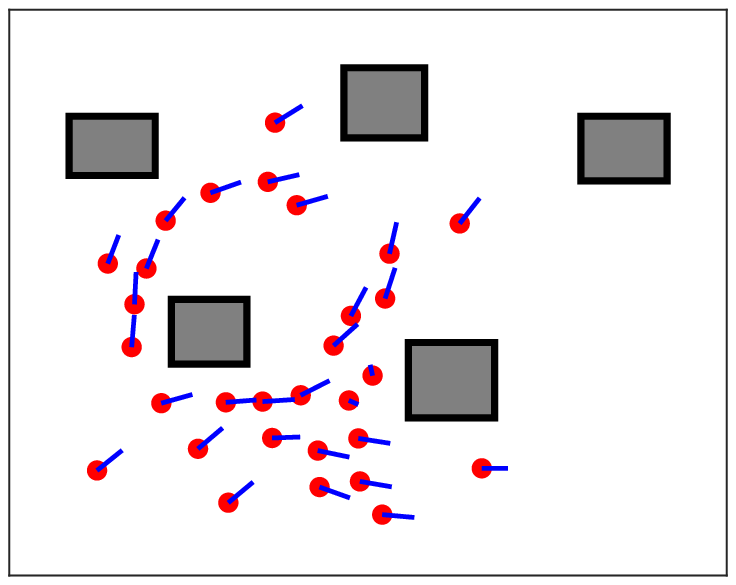}}\hfill
\subfloat[Predator avoidance]{\includegraphics[width=.24\textwidth]{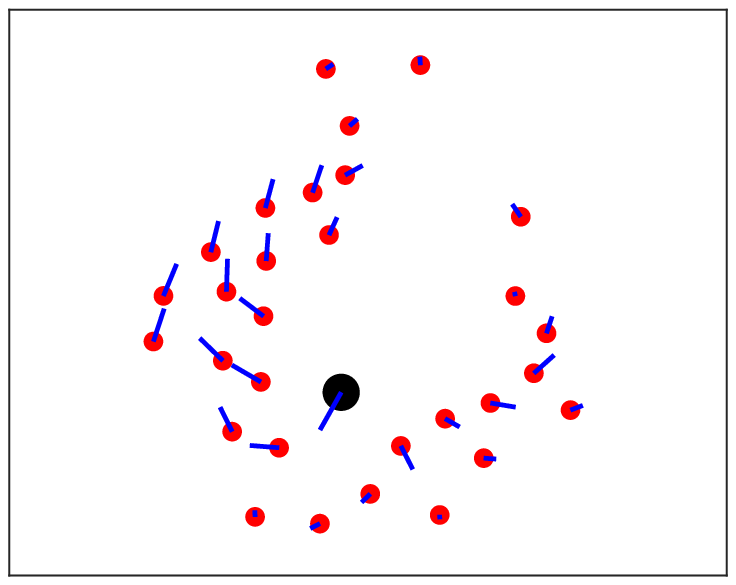}}\hfill
\subfloat[Target seeking]{\includegraphics[width=.24\textwidth]{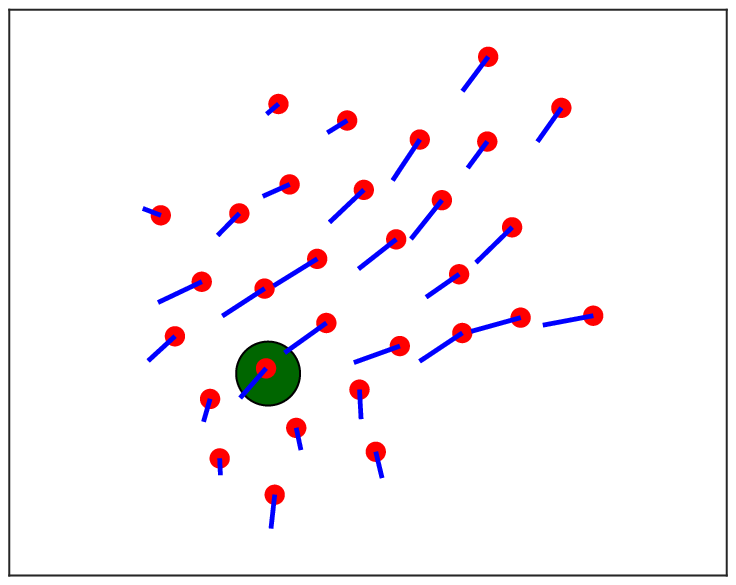}}
\caption{Snapshots of flocking behaviors }\label{fig:snapshots}
\end{figure}
While the simple cost function in Eqs.~(\ref{eq:central_mpc}) and~(\ref{eq_dmpc}) for basic flocking guarantees that in the steady state the agents are well separated, it does not guarantee collision avoidance. 
Additional goals such as collision avoidance and obstacle avoidance are added to the MPC problem as \emph{minimum distance constraints}. The constrained MPC is recast as an equivalent unconstrained MPC problem~\cite{Kerrigan2000softconstraints} by converting the constraints to a $\emph{penalty term}$, using the theory of exact penalty functions~\cite{Fletcher1987}. The weighted penalty term, derived based on the magnitudes of the constraint violations, is added to the MPC cost function.

Consider a generic nonlinear constrained optimization problem, as given in Eq.~(\ref{eq:opt_const_a}). It can be recast as the unconstrained optimization problem given by Eq.~(\ref{eq:opt_const_b}):

\begin{subequations}
\label{eq:opt_const}
\begin{align}
    &\underset{\theta}{min} \ V(\theta), \ \mathrm{subject  \ to \ }  c(\theta) \preceq 0, \label{eq:opt_const_a} \\
\qquad
    &\underset{\theta}{min} \ V(\theta) + \rho \left \|  c(\theta)^+\right \| \label{eq:opt_const_b}
\end{align}
\end{subequations}

where $\rho$ is the weight of the penalty term, and $c(\theta)^+$ is a vector containing the magnitudes of the constraint violations. For a constraint $c_i$, the magnitude of the constraint violation $c_i^+$ is equal to $max(c_i, 0)$. According to Theorem~1 of~\cite{Kerrigan2000softconstraints}, the solution of  problem~(\ref{eq:opt_const_a}) is equal to the one of problem~(\ref{eq:opt_const_b}) if $\rho > \| \lambda^* \|_D$ and $c(\theta^*) \preceq 0$.  Here $\lambda^*$ is the optimal Lagrange multiplier vector, $\| \cdot \|_D$ is the dual norm, and $\theta^*$ is the optimal solution.

We introduce cost function terms for collision avoidance, obstacle avoidance, target seeking, and predator avoidance.  Objectives can be combined by including the corresponding terms in the cost function. 

\paragraph{Cost Function Term for Collision Avoidance.}
For collision avoidance in a flock of $n$ agents, $n(n\,{-}\,1)\,{/}\,{2}$ pairwise constraints of the form $d_{min} - \|p_{ij}\| \leq 0 $ for all $i \neq j$ are applied to the MPC optimization problem. The pairwise constraints ensure that no two agents are closer than a distance of $d_{min}$ apart from each other. The constraints are converted into a penalty term, that is the 2-norm of the magnitude of constraint violations. For a pair of agents $i, j$, the magnitude of the constraint violation is equal to $max(d_{min} - \|p_{ij}\|, 0)$. Hence, the cost-function term for collision avoidance is $J_{CA} \left(\mathbf{p}\right) = \|c_{ca}(p)^+ \|$, where $c_{ca}(p)^+$ is the vector of the magnitude of collision-avoidance constraint violations. 

\paragraph{Cost Function Term for Obstacle Avoidance.}
For a flock of size $n$ and an obstacle field containing $m$ obstacles, $m\,n$ constraints of the form $d_{min} - \|p_{i} - o_{k}\| \leq 0$ are applied to the optimization problem, where $\|p_{i} - o_{k}\|$ is the distance between agent $i$ and the closest point on obstacle $k$.  For an agent-obstacle pair ($i, k$), the magnitude of the constraint violation is equal to $max(d_{min} - \|p_i - o_k\|, 0)$. Hence, the cost-function term for collision avoidance is, $J_{OA} \left(\mathbf{p}, \mathbf{o} \right) = \|c_{oa}(\mathbf{p}, \mathbf{o})^+ \|$, where $c_{oa}(\mathbf{p}, \mathbf{o})^+$ is the vector of the magnitudes of obstacle-avoidance constraint violations, and $\mathbf{o}$ is the set of points on obstacle boundaries.

\paragraph{Cost Function Term for Target Seeking.}
This term is the average of the squared distance between the agents and the target. Let $g$ denote the position of the fixed target.  Then the term is as defined as $    J_{TS}(\mathbf{p}) = \frac{1}{n} \sum_{i \in \mathcal{A} } \| p_i - g \|^2 $.

\paragraph{Cost Function Term for Predator Avoidance.}
We introduce one predator, which is more agile than the flocking agents, with a maximum speed and maximum acceleration a factor of $f_p > 1$ times higher than $\bar{v}$ and $\bar{a}$, respectively.  Apart from being more agile, the predator has the same dynamics as the agents, given by Eq.~(\ref{motioneq}). The control law for the predator consist of a single rule that causes it seek the centroid of the flock with maximum acceleration.

For a flock of $n$ agents and one predator, $n$ constraints of the form $d^{pred}_{min} - \|p_{i} - p_{pred}\| \leq 0$ are applied to the optimization problem, where $\|p_{i} - p_{pred}\|$ is the distance between agent $i$ and the predator, and $d^{pred}_{min}$ is the desired minimum distance. The magnitude of constraint violation for an agent $i$ is $max(d^{pred}_{min} - \|p_{i} - p_{pred}\|, 0)$. The cost function term is $ J_{PA} \left(\mathbf{p}, p_{pred}\right) = \|c_p(p, p_{pred})^+ \|$, where $c_p(p, p_{pred})^+$ is the vector of the magnitudes of predator-avoidance constraint violations.

\paragraph{Cost Function terms used in the experiments.}
The cost functions for our experiments are weighted sums of the cost function terms introduced above. We refer to the first term of Eq.~(\ref{eq_cmpc}) as $J_{cohes}(\mathbf{p})$ and the second as $J_{sep}(\mathbf{p})$. 
If in an experiment collision avoidance is added as penalty term, then the separation term is omitted due to redundancy. If multiple objectives expressed as constraints are used, the constraint violations are collected in one penalty term with weight $\rho$.  We use following cost functions $J_1$, $J_2$, $J_3$, and $J_4$ for experiments with basic flocking, collision avoidance, obstacle avoidance with target seeking, and predator avoidance, respectively.
\begin{subequations}
    \label{eq:cst_fnc_4}
    \begin{align}
    \begin{split}
    J_1(\mathbf{p}) = J_{cohes}(\mathbf{p}) + \omega \cdot J_{sep}(\mathbf{p})\label{eq:subflocking}
    \end{split}\\
    \begin{split}
    J_2(\mathbf{p}) = J_{cohes}(\mathbf{p} )+ \rho \cdot J_{CA}(\mathbf{p} ) \label{eq:sub_ca}
    \end{split}\\
    \begin{split}
    J_3(\mathbf{p}, \mathbf{o}) &= J_{cohes}(\mathbf{p} )+ \omega_t \cdot J_{TS}(\mathbf{p}) \\ &+ \rho \cdot \sqrt{J_{CA}(\mathbf{p})^2 + J_{OA}(\mathbf{p}, \mathbf{o})^2} \label{eq:sub_oa}
    \end{split}\\
    \begin{split}
    J_4(\mathbf{p}, p_{pred}) &= J_{cohes}(\mathbf{p} ) \\  &+ \rho \cdot \sqrt{J_{CA}(\mathbf{p})^2 +   J_{PA}(\mathbf{p}, p_{pred})^2} \label{eq:sub_pa}   
    \end{split}
    \end{align}
\end{subequations}
where $\omega_t$ is the weight of the target-seeking term. 

\subsection{Neural Network Architectures}

We consider two main neural network (NN) architectures. The first is a class of recurrent neural networks called \emph{Long Short Term Memory} (LSTM)~\cite{lstm}. LSTM have been shown to perform well in motion planning~\cite{Everettlstm} and obstacle-field navigation~\cite{chenlstm}. Our motivation for using LSTM was to exploit the temporal nature of the trajectory data, as LSTMs employ memory cells well suited for handling temporal data. 
The second class of NN we use is \emph{Deep Neural Network} (DNN). The performance of the DNC controllers we obtain strongly depends upon the chosen NN architecture. We refer to the resulting DNC controllers as DNC-LSTM and DNC-DNN, respectively.

\subsection{Training Distributed Flocking Controllers}
 Our objective is to learn basic flocking, collision avoidance, obstacle avoidance with target seeking, and predator avoidance.  The last two implicitly also include collision avoidance.
For each of these tasks, our methodology is to train a DNC using the trajectory data obtained from the CMPC. CMPC is run with a neighborhood size of $N=5$, starting from 100 random initial states, producing 100 trajectories, each with a duration of 100 time units.  We learn a single DNC from the state-action pairs of all $n$ agents. This yields a symmetric distributed controller, which we use for each agent during evaluation. 

\paragraph{Basic Flocking.} 
Trajectory data for basic flocking is generated using the cost function in Eq.~(\ref{eq_cmpc}). The input to the NN is the position and velocity of the agent along with the positions and velocities of its $N$-nearest neighbors. We will refer to the agent (DNC) being learned as $\mathcal{A}_0$.  Since we use neighborhood size $N=5$, the input to the NN is of the form $[p^x_0 \,\, p^y_0 \,\, v^x_0 \,\, v^y_0 \,\, p^x_1 \,\, p^y_1 \,\, v^x_1 \,\, v^y_1\cdot \cdot \cdot \cdot \, p^x_5 \,\, p^y_5 \,\, v^x_5 \, v^y_5]^T$, where $p^x_0$, $p^y_0$ are the position coordinates, and  $v^x_0$, $v^y_0$ velocity coordinates for the agent $\mathcal{A}_0$. Similarly, $p^x_{1...5}$, $p^y_{1...5}$, $v^x_{1...5}$ and $v^y_{1...5}$ are the position and velocity vectors of its neighbors.  Since this vector has 24 components, the input to the NN consists of 24 features.

\paragraph{Collision Avoidance.}
The CMPC cost function for collision avoidance is given in Eq. (\ref{eq:sub_ca}).
The input to the NN is the same as for basic flocking.

\paragraph{Obstacle Avoidance with Target Seeking.}
For obstacle avoidance with target seeking (and collision avoidance), we use CMPC with the cost function in Eq.~(\ref{eq:sub_oa}). The target is located behind the obstacles, forcing the agents to move through the obstacle field. 
For this task, the input to the NN is given by the positions and velocities of agent $\mathcal{A}_0$ along with its $N$-nearest neighbors as well as the position of the closest point on the obstacle from agent $\mathcal{A}_0$ and its $N$-nearest neighbors and the target location of the flock. The input to the NN consists of the 38 features $[p^x_0 \,\, p^y_0 \,\, v^x_0 \,\, v^y_0 \,\, o^x_0 \,\, o^y_0 \cdot \cdot \cdot \cdot \, p^x_5 \,\, p^y_5 \,\, v^x_5 \, v^y_5 \,\, o^x_5 \,\, o^y_5 \,\,g^x \,\,g^y]^T$, where $o_0^x$ , $o_0^y$ is the closest point to agent $\mathcal{A}_0$ on any obstacle; $o_{1...5}^x$ , $o_{1...5}^y$ give the closest point on any obstacle for the 5 neighboring agents, and $g^x$, $g^y$ is the target location. 

\paragraph{Predator Avoidance.}
The CMPC cost function for predator avoidance (with collision avoidance) is given in Eq. (\ref{eq:sub_pa}). The position, velocity and the acceleration of the predator are denoted by $p_{pred}$ , $v_{pred}$ , $a_{pred}$, respectively.   We take $f_p = 1.25$, hence $\bar v_{pred} = 1.25 \, \bar v$ and $\bar a_{pred} = 1.25 \, \bar a$. 
The input features to the neural network are the positions and velocities of agent $\mathcal{A}_0$ along with its $N$-nearest neighbors and the position and velocity of the predator. The input with 28 features has the form $[p^x_0 \,\, p^y_0 \,\, v^x_0 \,\, v^y_0 \cdot \cdot \cdot \cdot \, p^x_5 \,\, p^y_5 \,\, v^x_5 \, v^y_5\,\,  p^x_{pred} \,\,  p^y_{pred} \,\,  v^x_{pred} \,\,  v^y_{pred}]^T$.

\section{Experimental Evaluation}
\label{sec:eval}

We conducted an extensive performance evaluation of Neural Flocking, taking into account various control objectives: basic flocking, collision avoidance, obstacle avoidance with target seeking, and predator avoidance.  As illustrated in Fig.~\ref{fig:architecture}, this involved running CMPC to generate the training data for the distributed neural controllers, whose performance was then compared to the DMPC controllers.  We also showed that the learned DNC flocking controllers generalize the training data in two important ways: they achieve successful collision-free flocking in flocks larger than those used in the training data, and they achieve obstacle avoidance in obstacles fields having a larger number of obstacles than what was present in the training data.  We include as supplementary material multiple videos depicting the quality of learning for these controllers.  We consider different NN architectures, each with $O(10^4)$ training parameters.  

The CMPC and DMPC problems are solved using gradient-descent optimization. In the training phase, the size of the flock is $n = 30$. For obstacle-avoidance experiments, we use 5 obstacles. The simulation time is $100$, $dt\,{=}\,0.1$ time units, $\eta=3$, $\Bar{v}\,{=}\,2.0$ and $\Bar{a}\,{=}\,1.5$. As reported in~\cite{Mehmood2018}, the weight $\omega$ of the separation term in DMPC and CMPC is $30$ and $2000$, respectively. We use $d_{min}=2$ and $d^{pred}_{min}=4$; a higher value is used for $d^{pred}_{min}$ due to the agility of the predator.  The weight of the penalty term is $\rho\,{=}\,100,000$.  For the initial configuration, the positions and velocities are uniformly sampled from $[-15, 15]^2$ and $[0, 1]^2$, respectively. We ensure that the initial configuration is \emph{recoverable}; i.e., no two agents are so close to each other that they cannot avoid a collision when resorting to maximal acceleration. The predator starts at rest from a fixed location at a distance of 50 from the flock center. 

For training, we considered 30 agents and 100 trajectories per agent, each trajectory 334 time steps in length.  This yielded a total of 1,002,000 training samples. We use two variants of neural nets, LSTM and DNN, to learn DNCs. For LSTM, we use 2 hidden layers with 34 cells per hidden layer, with a sigmoid activation function. For the DNN version, we use 5 hidden layers with 64 neurons per hidden layer, with a sigmoid activation function. We chose the number of hidden layers and neurons such that the numbers of trainable parameters are comparable for both classes of NNs.

The Adam optimizer~\cite{adamopt} was used with the following settings: $lr\,{=}\,10^{-4}$, $\beta_1\,{=}\,0.9$, $\beta_2\,{=}\,0.999$, $\epsilon\,{=}\,10^{-8}$. The number of epochs used for training is 10,000 and the batch size is 500. For measuring training loss, we use the mean-squared error metric. For both basic flocking and collision-avoidance, we give the neural network an input vector with 24 features. The number of trainable parameters for these two control objectives for the DNN and LSTM configurations are 18,370 and 18,644, respectively.  For obstacle-avoidance and target-seeking, we give the neural network an input vector with 38 features. The number of trainable parameters for the DNN and LSTM are 19,138 and 20,324, respectively. Finally, the predator-avoidance control objective has 28 features as the input to the neural network; the resulting number of trainable parameters for the DNN and LSTM architectures are 19,266 and 20,604, respectively. For training the neural networks, we use Keras~\cite{chollet2015keras}, which is a high-level neural network API written in Python and capable of running on top of TensorFlow.

To test the learned DNCs, we generated 100 simulations (runs) for each of the desired control objectives: basic flocking, flocking with collision avoidance, flocking with obstacle avoidance and target seeking, and flocking with predator avoidance.  The results presented in Tables~\ref{table:flock}-\ref{table:oa+pa}, were obtained using the same number of agents and obstacles and the same predator as in the training phase. We also ran tests that show DNC controllers can achieve collision-free flocking with obstacle avoidance where the numbers of agents and obstacles are greater than those used during training.  The Supplemental Material includes videos demonstrating flocking with 35 agents and obstacle avoidance with 10 obstacles.

We use flock diameter and velocity convergence~\cite{zhang2015model} as performance metrics for flocking behavior. At any time step, the \emph{flock diameter} $D(\mathbf{p}) =\max_{(i,j) \in \mathcal{A}} \Vert p_{ij} \Vert$ is the largest distance between any two agents in the flock. The velocity convergence \emph{VC}$(\mathbf{v})= (1/n)\left (\sum_{i \in \mathcal{A}}  \Vert v_i - (\sum_{j=1}^n v_j)/n  \Vert^2 \right )$ is the average of the magnitude of the discrepancy between the velocities of agents and the flock's average velocity.  For both metrics, lower values are better, indicating a dense and coherent flock.  A successful flocking controller should also ensure that both values eventually stabilize.

For collision avoidance, obstacle avoidance, and predator avoidance, collision rates are used as a performance metric.  An inter-agent collision (IC) occurs when the distance between two agents at any point in time is less than $d_{min}$.  An obstacle-agent collision (OC) occurs when the distance between an agent and the closest point on any obstacle is less than $d_{min}$.  A predator-agent collision (PC) occurs when the distance between an agent and the predator is less than $d^{pred}_{min}$.  The collision counts reported below are the total numbers of collisions of each type in the 100 test trajectories. The collision rate is the the number of states in those trajectories in which a collision occurs divided by the total number of states in those trajectories. 

\begin{table}[t!]
  \caption{Performance comparison for basic flocking} 
  \label{table:flock}
  \centering
  \resizebox{\linewidth}{!}{%
  \begin{tabular}{ccccc}
    \toprule
    Models    & Average     & SD of     & Velocity     & SD of Velocity\\
              & Converged Diameter & Converged Diameter & Convergence & Convergence\\
    \midrule
    DNC-DNN   & 22.8138  & 2.0137  & 0.0406  & 0.0037 \\
    DNC-LSTM  & 23.7629  & 2.8272  & 0.0435  & 0.0041 \\
    DMPC      & 21.1231  & 2.5358  & 0.0376  & 0.0036 \\
    CMPC      & 22.0111  & 2.6494  & 0.0303  & 0.0012 \\
    \bottomrule
  \end{tabular}}
\end{table}

\begin{table}
  \caption{Performance comparison for flocking with collision avoidance}
  \label{table:ca}
  \centering
  \begin{tabular}{ccccc}
    \toprule
    Models    & Average     & SD of     & Velocity  & SD of \\
              & Converged & Converged & Convergence & Velocity\\
              & Diameter & Diameter &  & Convergence\\
    \midrule
    DNC-DNN   & 22.3521  & 2.0362  & 0.1501  & 0.0180  \\
    DNC-LSTM  & 23.2676  & 2.2193  & 0.1638  & 0.0211 \\
    DMPC      & 61.7392  & 34.8767  & 0.2499  & 0.1941 \\
    CMPC      & 21.4441  & 1.8788  & 0.1449  & 0.0258 \\
    \bottomrule
  \end{tabular}
\end{table}

\begin{figure}[t]
\centering
\includegraphics[width=0.5\textwidth, trim=0cm 7.05cm 0cm .69cm, clip]{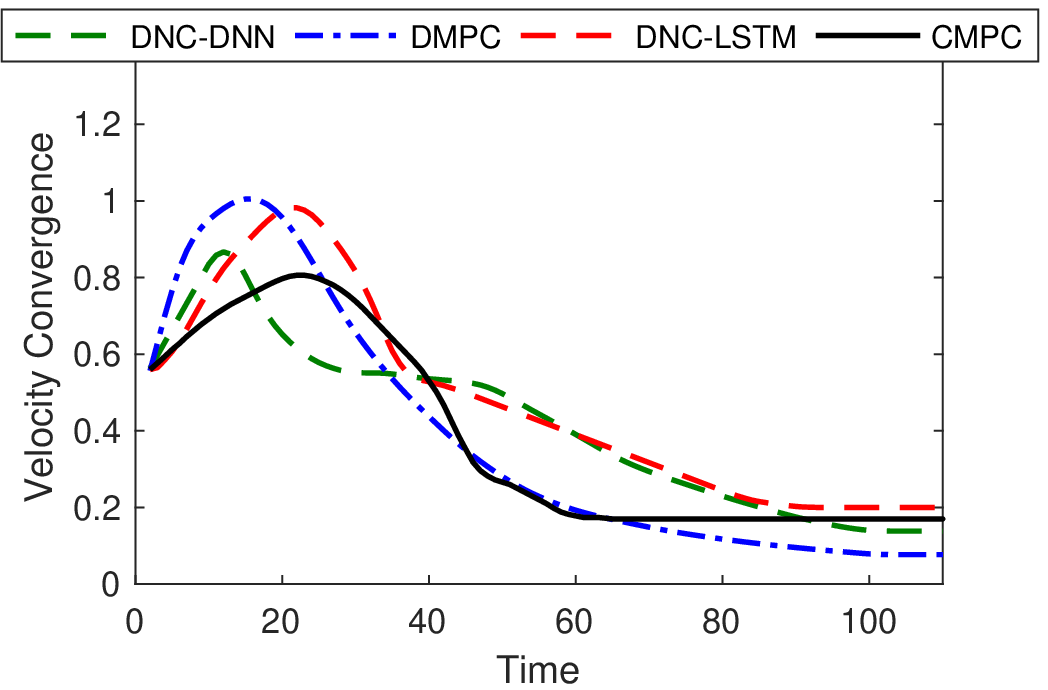}\\ \vspace*{10pt}
\subfloat[Flock diameter for basic flocking]{\includegraphics[width=.24\textwidth]{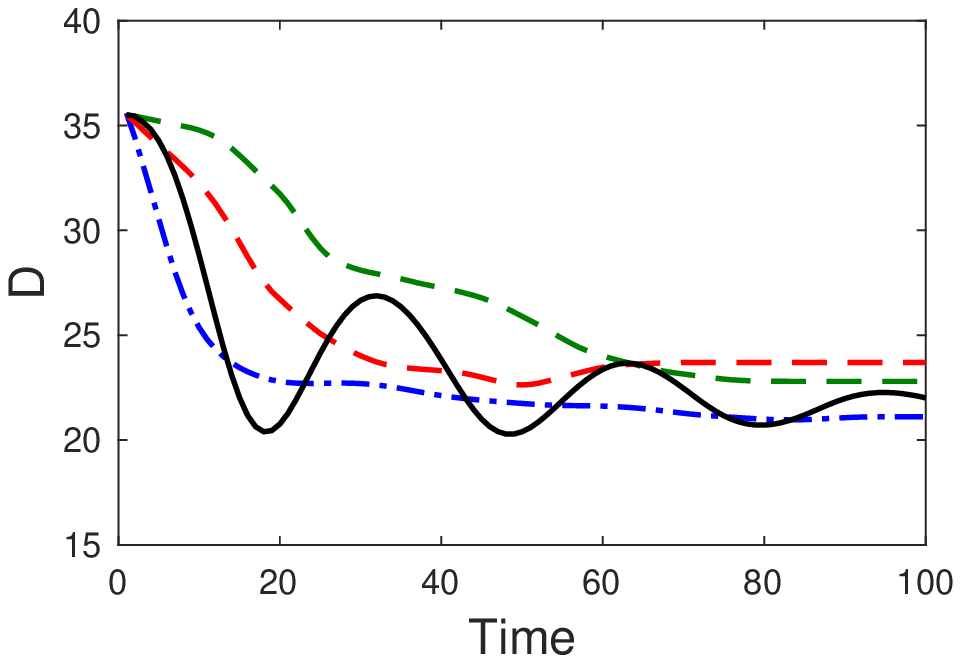}}\hfill
\subfloat[Flock diameter for collision avoidance]{\includegraphics[width=.24\textwidth]{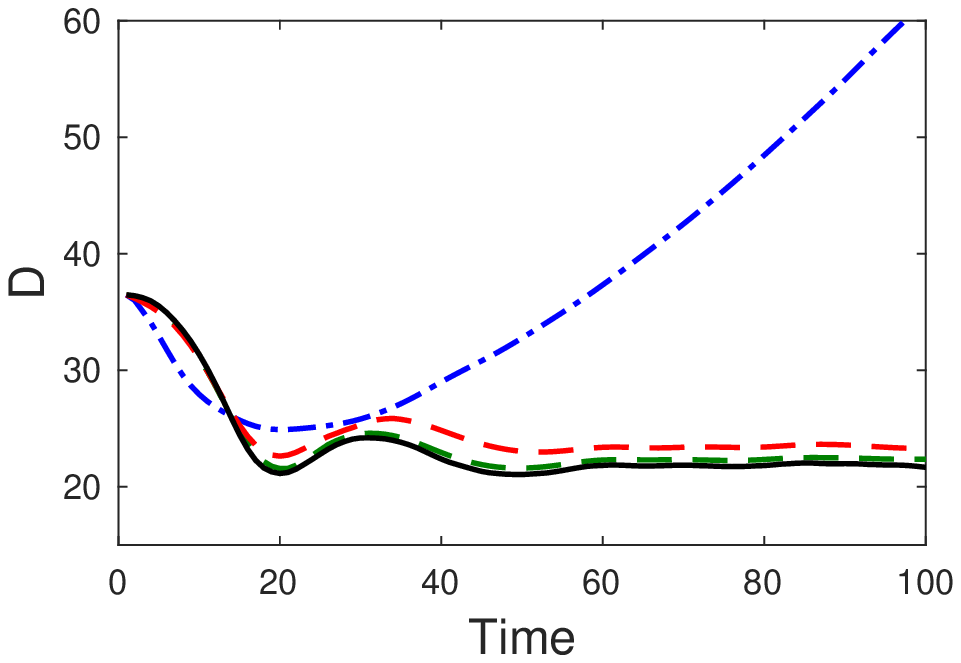}}\hfill
\subfloat[Velocity convergence for basic flocking]{\includegraphics[width=.24\textwidth]{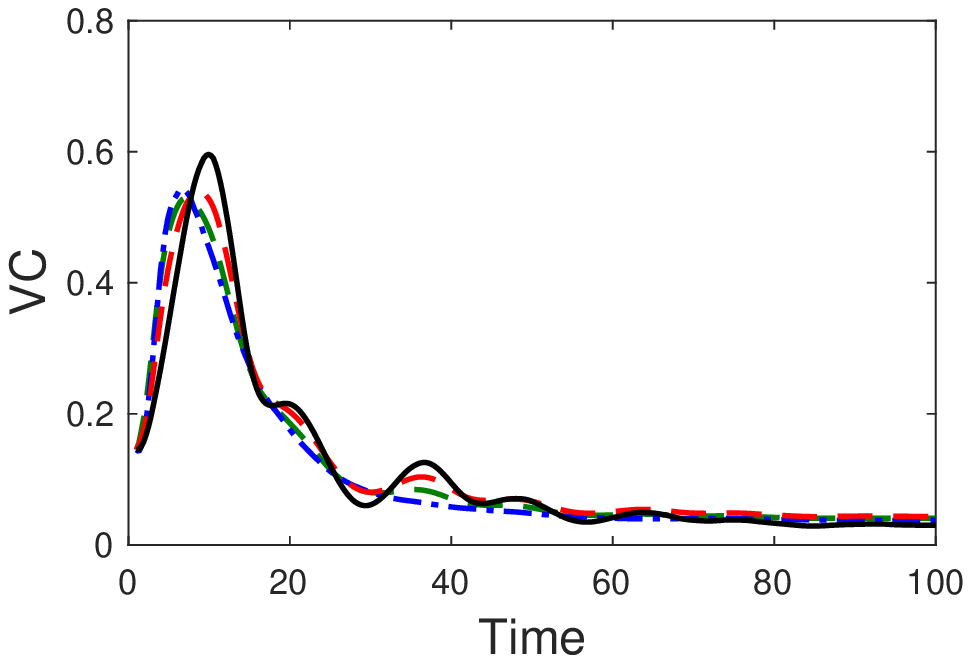}}\hfill
\subfloat[Velocity convergence for collision avoidance]{\includegraphics[width=.24\textwidth]{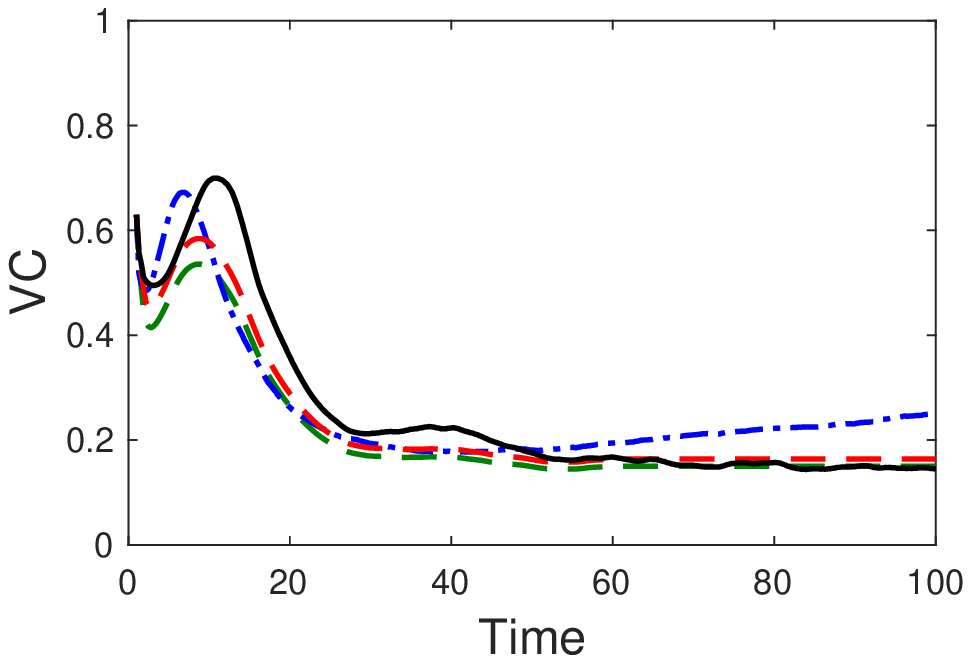}}
\caption{Comparison of performance measures for basic flocking and collision avoidance, averaged over a 100 runs for each flocking controller.}\label{fig:plots}
\end{figure}

We calculate the average converged diameter by averaging the flock diameter in the final time step of the simulation over the 100 runs. Similarly, the standard deviation of the velocity convergence is obtained from the last time step of all 100 runs. Table~\ref{table:flock} shows the performance of the DNC variants against the MPC controllers with respect to basic flocking for 30 agents. Although the DMPC performance is better than DNC-DNN and DNC-LSTM, the difference is marginal. Table~\ref{table:ca} presents the collision-avoidance results for 30 agents.  Both DNC-DNN and DNC-LSTM outperform the DMPC controller in terms of flock diameter and velocity convergence. As seen in Fig.~(\ref{fig:plots}b), DMPC does poorly for collision avoidance due to flock fragmentation; this leads to an increase in flock diameter. Although the DNCs are also distributed, they do not encounter flock fragmentation. This is likely because they are trained using CMPC-generated data, and the CMPC controller has a flock-wide view of the system. Thus, the DNCs are able to learn patterns in the state of their neighbors that help them flock better.

Table~\ref{table:oa+pa} shows the collision rates for the DNC controllers and DMPC.   An ideal controller should produce no collisions.  The DNCs achieve zero inter-agent collisions.  Furthermore, for all three types of collisions, the DNCs achieve significantly fewer collisions than the DMPC. 

\begin{table}[t!]
  \caption{Performance comparison for collision, obstacle and predator avoidance}
  \label{table:oa+pa}
  \centering
  \begin{tabular}{ccccccc}
    \toprule
      Models  & IC   & IC   & OC     & OC   & PC     & PC  \\
       & Count & Rate & Count & Rate & Count & Rate \\
    \midrule
    DNC-DNN    & 0  & 0\%  & 311  & 0.93\%  & 1621  & 4.85\%\\
    DNC-LSTM   & 0  & 0\% & 338  & 1.01\%  & 1737  & 5.20\%\\
    DMPC    & 17  & 0.05\% & 569  & 1.70\%  & 2130  & 6.38 \%\\
    \bottomrule
  \end{tabular}
\end{table}

Based on the experimental results, we make two main observations about learnt distributed controllers and distributed model predictive controllers. First, 
an important advantage of DNCs over DMPCs is that they are much faster.  Executing a DNC requires a modest number of arithmetic operations, whereas executing an MPC requires simulation of a model and controller over the prediction horizon.  In our experiments, on average, the CMPC and DMPC take 10 msec and 57 msec of CPU time, respectively, whereas DNC-DNN and DNC-LSTM take only 1.6 msec and 1.8 msec, respectively.
Second, DMPCs are well suited for certain classes of cost functions, and do not perform so well on others. The general intuition is as follows: if the cost function does not involve additive terms that couple the states of different agents, that is, the cost is separable, then the individual agents are essentially decoupled and distributed MPC approaches will work well. If the cost involves some limited form of coupling, for example, additive terms in the cost function only involve states of neighbors (and no coupling between non-neighbors), then again distributed MPC {\em{may}} perform well. If cost function also couples together non-neighbors, then distributed MPC is unlikely to perform well without some coordination~\cite{dunbar2004,dunbar2006}. However, the syntactic properties of the cost function that make it more amenable to a distributed MPC approach are often only sufficient, but not necessary, for distributed MPC to perform well. In general, it is not easy to determine whether a particular cost function is a good match for a distributed MPC approach or not. In the case of flocking, syntactically the cost function involves coupling among all agents. However, DMPC's performance is good for basic flocking, but it deteriorates soon when the cost for basic flocking is enhanced with other terms. Our approach of training DNC using CMPC is more robust along this dimension. In the results presented above, while DMPC's performance is uneven, DNC continues to perform well as we add nonlinear terms for collision avoidance and terms for obstacle and predator avoidance. In some sense, DNC appears to implicitly learn some aspects about the required coordination whenever it is needed for a new cost function.

\section{Evaluating Generalization to Complex Plant Models}
\label{sec:TL}

We have successfully demonstrated that a distributed controller can be effectively synthesized from training samples generated from a centralized controller. In fact, the learnt distributed controller was shown to generalize as well as, if not better than, the distributed MPC controller. We evaluated the two different approaches for obtaining distributed controllers by varying the number of agents, number of obstacles, and changing the objective cost function.  In this section, we ask the following question: does a distributed neural controller (DNC) generalize as well as the centralized MPC controller (CMPC) when we change the plant model to a richer plant model?

To answer the above question, we will train the distributed neural controller from data generated using CMPC on a simplified model and then see how well the learnt DNC controller performs when the plant is a collection of quadrotors (in simulation).
Since the quadrotor model uses positions in 3-dimensions, we first extend our 2d point model (Equation~(1)) into a 3d model (in the obvious way by adding a third dimension with the same dynamics). Following the same methodology as described earlier for 2-dimensions, we train a distributed neural controller for 3-dimensions. 

\subsection{Training DNC for 3D point model}
We train the DNC for the basic flocking task. The neural network takes as input a 36-dimensional vector of the form  $[p^x_0 \,\, p^y_0 \,\,p^z_0 \,\, v^x_0 \,\, v^y_0 \,\,v^z_0 \,\, p^x_1 \,\, p^y_1 \,\,p^z_1 \,\, v^x_1 \,\, v^y_1 \,\,v^z_1 \,\,\cdot \cdot \cdot \cdot \, p^x_5 \,\, p^y_5 \,\,p^z_5 \,\, v^x_5 \, v^y_5 \,\,v^z_5 \,\,]^T$, where $p^x_0$, $p^y_0$, $p^z_0$ are the position coordinates, and  $v^x_0$, $v^y_0$, $v^z_0$ are velocity coordinates for the agent $\mathcal{A}_0$ (running the controller). Similarly, $p^x_{1...5}$, $p^y_{1...5}$, $p^z_{1...5}$, $v^x_{1...5}$, $v^y_{1...5}$ and $v^z_{1...5}$ are the position and velocity vectors of $\mathcal{A}_0$'s neighbors respectively. This is similar to the inputs used by the NN for basic flocking in our 2-dimensional representation, with the only difference being the addition of a third dimension, thus making the number of input features to be 36.

For training the DNN with CMPC data obtained using the 3D point-model, we consider 20 agents and 400 trajectories per agent, each trajectory 400 time steps in length.  This yielded a total of 3,200,000 training samples. The DNN uses 5 hidden layers with 84 neurons per hidden layer, with a ReLU activation function. The number of trainable parameters and training epochs are 31,923 and 10,000 respectively. The parameters for the Adam optimizer are same as those specified in Section~\ref{sec:eval}.

We want to see how DNC and CMPC generalize from controlling a flock of points to controlling a flock of quadrotors. So, we now describe the dynamical model of the quadrotor.

\subsection{Quadrotor Dynamics}
\begin{figure}[t]
    \centering
    \includegraphics[width=8cm]{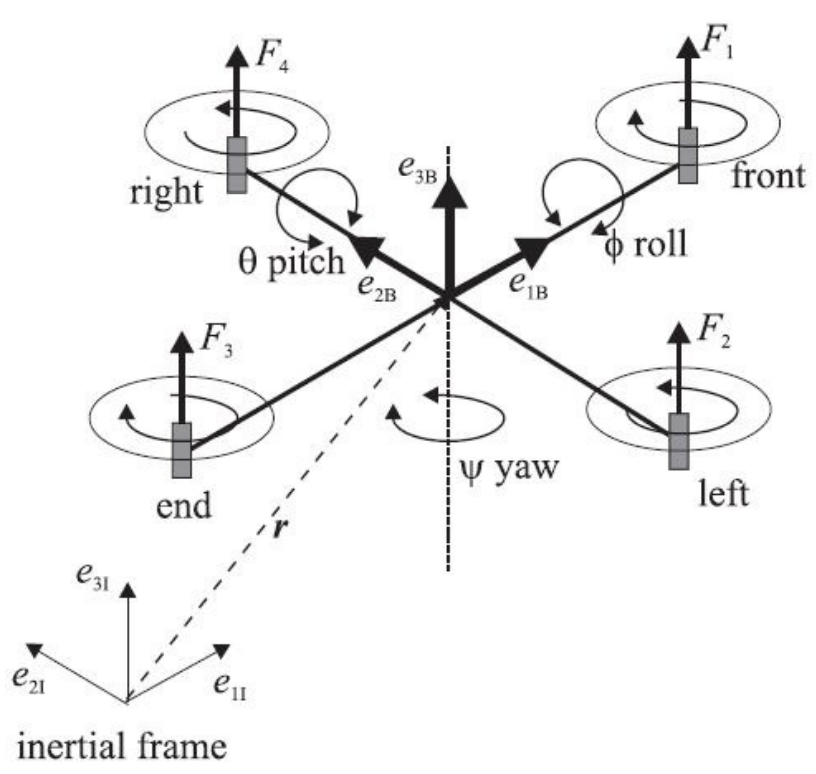}
    \caption{ Coordinate frame of the quadrotor}
    \label{fig:quad}
\end{figure}

The equations of motion of the quadrotor are  derived using two coordinate frames as shown in Fig.~\ref{fig:quad}. Model derivation is based on the assumptions that the quadrotor is a rigid body, it has a symmetrical structure and the center of gravity and the body frame origin are coincided.

The orientation of quadrotor is specified by euler angles $\phi, \theta, \psi$ which denote the roll, pitch and yaw, respectively. $r^T = [x \,\, y \,\, z]$ is the position of quadrotor in inertial frame. The dynamical equations of quadrotor can be written in following form~\cite{Bouabdallah2007}:

\begin{align} \label{eq:quad}
\begin{split}
    \ddot x &= (\cos \phi \sin \theta \cos \psi + \sin \phi \sin \psi) \frac{u_1}{m} \\
    \ddot y &= (\cos \phi \sin \theta \sin \psi - \sin \phi \cos \psi) \frac{u_1}{m} \\
    \ddot z &= -g + \cos \phi \cos \theta \frac{u_1}{m} \\
    \ddot \phi &= \dot \theta \dot \psi \frac{I_{yy}-I_{zz}}{I_{xx}} - \frac{J_r}{I_{xx}} \dot \theta \Omega_r + \frac{u_2}{I_{xx}} \\
    \ddot \theta &= \dot \psi \dot \phi \frac{I_{zz}-I_{xx}}{I_{yy}} + \frac{J_r}{I_{yy}} \dot \phi \Omega_r + \frac{u_3}{I_{yy}} \\
    \ddot \psi &= \dot \phi \dot \theta \frac{I_{xx}-I_{yy}}{I_{zz}} + \frac{u_4}{I_{zz}}
\end{split}
\end{align}
where, $m$ is the total mass of the quadrotor, $g$ denotes the acceleration due to gravity, $I_{xx}$, $I_{yy}$ and $I_{zz}$ are inertias of the quadrotor and $J_r$ denotes the inertia of the propeller. In addition, $\Omega_r = \Omega_1 + \Omega_2 + \Omega_3 + \Omega_4 $ is the total angular speed of propellers ($\Omega_i$ is angular speed of $i$th rotor). $u_1$, $u_2$, $u_3$ and $u_4$ in Eq.~\ref{eq:quad} are the control inputs.

\begin{align} \label{eq:quadcontrol}
\begin{split}
    u_1 &= F_1 + F_2 + F_3 + F_4 \\
    u_2 &= L(F_4 - F_2) \\
    u_3 &= L(F_3 - F_1) \\
    u_4 &= -Q_1 + Q_2 - Q_3 + Q_4 \\
\end{split}
\end{align}

$F_i = b\Omega^2_i$ is the thrust produced by rotors which are perpendicular to $xy$ plane and $Q_i = d\Omega^2_i$ denotes drag moment of $i$th rotor. $b$ and $d$ are thrust and drag factors respectively, $L$ denotes horizontal distance between propeller center to center of gravity. 

The state and input vectors for the dynamic model of the quadrotor are given by $x^T = [x\,\,\dot x\,y\,\,\dot y\,z\,\,\dot z\,\,\phi \,\,\dot \phi \,\,\theta\,\,\dot \theta\,\,\psi \,\,\dot \psi ]$ and $u^T = [u_1 \,\, u_2 \,\, u_3 \,\, u_4]$ respectively. The parameters for the quadrotor are provided in Table~\ref{table:quadparams}.

Note that there is mismatch in the control inputs between the point model and the quadrotor model. The control inputs in the $3$d point model are the $3$d accelerations for each agent. Therefore, we designed a PID controller that takes as input a $3$d acceleration, and returns as output $[u_1,u_2,u_3,u_4]$.  The PID controller allows us to use the {\em{same}} outer loop controller on both the $3$d point model and the quadrotor model.

\subsection{From point to quadrotor model}

For testing the genaralization ability of DNC and CMPC, we perform four different kind of simulation runs:
\\
(1) DNC-point: We use the learnt DNC as a local controller on each agent to compute the $3$-dimensional acceleration vector for that agent. Each agent then uses the $3$d point model equations to update its position and velocity. The process repeats at each time step.
\\
(2) DNC-quad: We use the learnt DNC as a local controller on each agent to compute the $3$-dimensional acceleration for that agent. Each agent then uses the PID controller to compute $[u_1,u_2,u_3,u_4]$, which is then used in the quadrotor model to compute the next state of the agent.
\\
(3) CMPC-point: This is obtained by replacing the DNC controller in the DNC-point simulations by the CMPC controller.
\\
(4) CMPC-quad: This is obtained by replacing the DNC controller in the DNC-quad simulations by the CMPC controller. Note that the CMPC controller uses the point model internally when doing model-predictive control.

We note that the DNC controller used in DNC-point and DNC-quad simulations are identical. They are trained on the same data collected from trajectories generated by CMPC-point simulations (that is, simulations that use CMPC as the controller and point dynamics for the agents). Similarly, the CMPC controller used in CMPC-point and CMPC-quad simulations are identical.


\subsection{Results}

\def\dncp{{\mathtt{DNCpoint}}}
\def\dncq{{\mathtt{DNCquad}}}
\def\cmpcp{{\mathtt{CMPCpoint}}}
\def\cmpcq{{\mathtt{CMPCpoint}}}

We performed multiple runs of CMPC-point, CMPC-quad, DNC-point, and DNC-quad simulations.
Let $\cmpcp, \cmpcq, \dncp, \dncq$ denote the trajectories generated by the CMPC-point,
CMPC-quad, DNC-point, and DNC-quad simulations.

Let $\tau$ be a trajectory
$\langle s_0, s_1, s_2, \ldots\rangle$, where $s_t$ is the state of the flock at time $t$.
We lift the function $D$ that computes the {\em{flock diameter}} of a state to work over trajectories.
$$
D(\tau) = \langle D(s_0), D(s_1), D(s_2), \ldots\rangle
$$

Similarly, we lift the function $VC$ that computes the {\em{velocity convergence}} of a state to work over trajectories.
$$
VC(\tau) = \langle VC(s_0), VC(s_1), VC(s_2), \ldots \rangle
$$

\def\Tau{{\mathcal{T}}}

We now lift the functions $D$ and $VC$ to work over a set of trajectories $\Tau$:
$$
D(\Tau) = \{D(\tau) \mid \tau\in\Tau\},\quad
VC(\Tau) = \{VC(\tau) \mid \tau\in\Tau\}
$$
Let $avg$ compute a point-wise average of a set of vectors:
$$
avg(\vec{c}_1,\ldots,\vec{c}_k) = \sum_{i-1}^k \vec{c}_i / k
$$
where $\vec{c}_i$s are vectors of reals.

\begin{figure}[t]
\centering
\includegraphics[width=0.2\textwidth, trim=2.78cm 7.05cm 1.54cm .69cm, clip]{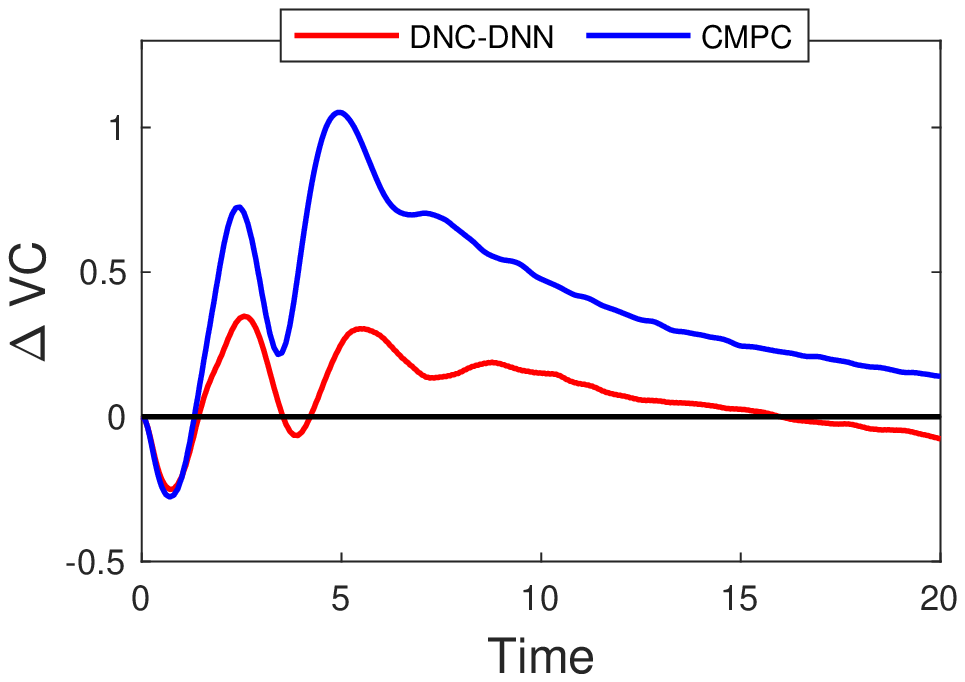}\\ \vspace*{10pt}
\subfloat[Difference in average diameter]{\includegraphics[width=.24\textwidth]{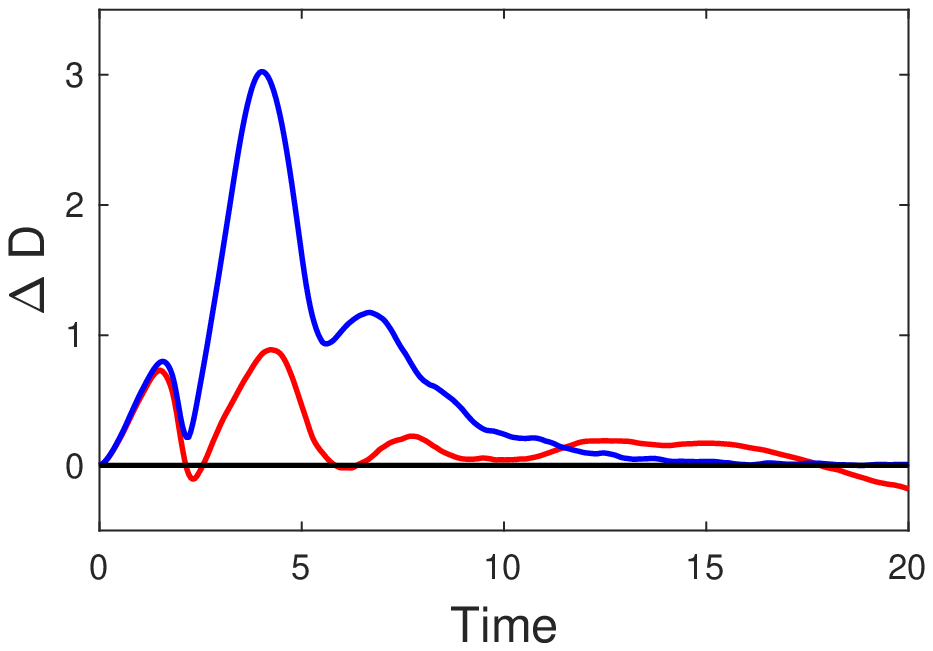}}\hfill
\subfloat[Difference in velocity convergence]{\includegraphics[width=.24\textwidth]{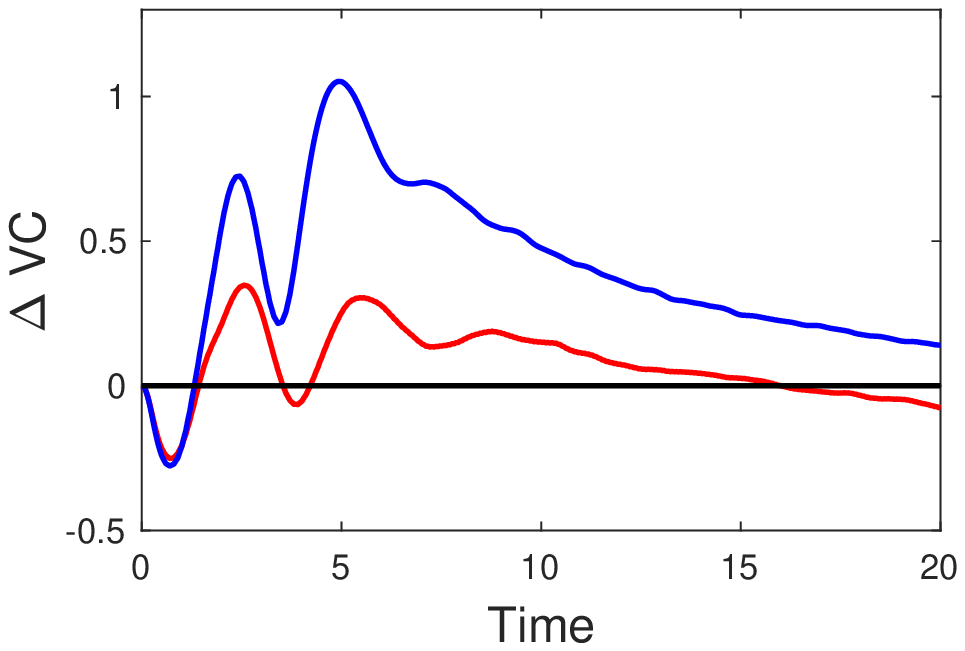}}
\caption{The difference between average performance measures for 3D point-model and quadrotor model (for both CMPC and DNC).}\label{fig:diffplots}
\end{figure}

Now, the average diameter of the flock over time when using simulations from DNC-point is
$avg(D(\dncp))$. The average diameter of the flock in the simulations from DNC-quad is 
$avg(D(\dncq))$. The difference $\Delta D = avg(D(\dncp)) - avg(D(\dncq))$ is a measure that shows how well DNC generalized when we replaced 3d point agents by quadrotor agents.
A value close to zero would indicate that DNC generalized well.  This difference, $\Delta D$, is shown as a red line in the plot in Figure~\ref{fig:diffplots}(a).
The same difference for the CMPC controller is shown as a blue line in that figure. Note that the red line remains closer to the $x$-axis compared to the blue line, which indicates DNC generalizes well to the quadrotor model.

Figure~\ref{fig:diffplots}(b) is obtained similarly, but for the velocity convergence metric (in place of the diameter). That is,
the red line in that plot shows $\Delta VC = avg(VC(\dncp)) - avg(VC(\dncq))$, and the blue line shows $avg(VC(\cmpcp)) - avg(VC(\cmpcq))$.
The results clearly indicate that the distributed neural controller generalizes as well as the CMPC controller when we replace the dynamics of the agents by a more complex quadrotor dynamics.


    

\begin{table}[t!]
\caption{Parameters of quadrotor}
  \label{table:quadparams}
  \centering
  \scalebox{1.1}{
  \begin{tabular}{ccc}
    \toprule
     Parameters &  Values  & Units \\
    \midrule
     $m$ & 0.650 & Kg \\
     $I_{xx}$ & 7.5e-3 & Kg.m$^2$ \\
     $I_{yy}$ & 7.5e-3 & Kg.m$^2$ \\
     $I_{zz}$ & 1.3e-2 & Kg.m$^2$ \\
     $J_r$ & 6e-5 & Kg.m$^2$ \\
     $L$ & 0.23 & m \\
     $b$ & 3.13e-5 & Ns$^2$ \\
     $d$ & 7.5e-7 & Nms$^2$ \\
    \bottomrule
  \end{tabular}}
\end{table}

\section{Related Work}
\label{sec:related}

The work in~\cite{bentley2018} synthesizes a flocking controller using multi-agent reinforcement learning (MARL) and natural evolution strategies (NES). The target model from which the system learns is Reynolds flocking model~\cite{REYNOLDS1987}. For training purposes, a list of metrics called \emph{entropy} are chosen which provide a measure of the collective behavior displayed by the target model.  As the authors of~\cite{bentley2018} observe, this technique does not quite work: although it consistently leads to agents forming recognizable patterns during simulation, agents self-organized into a cluster instead of flowing like a flock. The work of~\cite{La2015predavoid} combines reinforcement learning and flocking control for the purpose of predator avoidance, where the learning module determines safe spaces in which the flock can navigate to avoid predators. Their approach to predator avoidance, however, isn't distributed as it requires a majority consensus by the flock to determine its action to avoid predators. They also impose an $\alpha$-lattice structure~\cite{olfati2006flocking} on the flock to ensure predator avoidance. In contrast, our approach is geometry-agnostic and can achieve predator avoidance in a distributed manner.

The approach of~\cite{kahn2017RL} develops an uncertainty-aware reinforcement learning algorithm to estimate the probability of a mobile robot colliding with an obstacle in an unknown environment. Their approach is based on bootstrap neural network using dropouts, allowing it to process raw sensory inputs.  Similarly, a learning-based approach to robot navigation and obstacle avoidance is given in~\cite{PfeifferSNSC17}.  They train a model that maps sensor inputs and the target position to motion commands generated by the ROS~\cite{ros} navigation package. Our work in contrast considers obstacle avoidance (and other control objectives) in a multi-agent flocking scenario under the simplifying assumption of full state observation.  In~\cite{Godoy2016}, an approach based on Bayesian inference is proposed that allows an agent in a heterogeneous multi-agent environment to estimate the navigation model and goal of each of its neighbors.  It then uses this information to compute a plan that minimizes inter-agent collisions while allowing the agent to reach its goal.  Flocking formation is not considered in this paper.

Transfer learning algorithms aim to find an optimal transfer map between different models. A multi-robot transfer learning~\cite{Helwa2017} allows a quadrotor to use data generated by a second, similar quadrotor to improve its own behavior. In contrast our work doesn't just attempts to improve behavior but shows how our controller designed from a simpler model can generalize to a more complex model. \cite{Zhou2019}~uses online learning approach that enables knowledge transfer from the source quadrotor such that the target quadrotor achieves high-accuracy trajectory tracking with minimal data recollection and training. 

\section{Conclusions}
\label{sec:conclusions}
With the introduction of Neural Flocking (NF), we have shown in this paper how machine learning in the form of Supervised Learning can bring many benefits to the flocking problem.  As our  experimental evaluation confirms, the symmetric and distributed neural controllers we derive in this manner are capable of achieving a multitude of flocking-oriented flight objectives, including: flocking formation, inter-agent collision avoidance, obstacle avoidance, predator avoidance, and target seeking.  Moreover, NF controllers exhibit real-time performance and generalize the behavior seen in the training data to achieve these objectives in a significantly broader range of scenarios. Additionally, the neural controller exhibits the ability to generalize to richer plant models despite being trained upon a simpler point-based model.

For future work, we plan to investigate a distance-based notion of agent neighborhood as opposed to our current nearest-neighbors formulation.  
Furthermore, motivated by the quadcopter study of~\cite{ZhangKLA2016berkeley}, we will seek to combine MPC with reinforcement learning in the framework of guided policy search as an alternative solution technique for the NF problem.

\medskip
\small

\bibliographystyle{unsrt}
\bibliography{references}

\begin{thebibliography}{10}

\bibitem{REYNOLDS1987}
Craig~W. Reynolds.
\newblock Flocks, herds and schools: A distributed behavioral model.
\newblock {\em SIGGRAPH Comput. Graph.}, 21(4), August 1987.

\bibitem{Reynolds99}
Craig~W. Reynolds.
\newblock Steering behaviors for autonomous characters.
\newblock In {\em Proceedings of Game Developers Conference 1999}, pages
  763--782, 1999.

\bibitem{Mehmood2018}
Usama Mehmood, Nicola Paoletti, Dung Phan, Radu Grosu, Shan Lin, Scott~D.
  Stoller, Ashish Tiwari, Junxing Yang, and Scott~A. Smolka.
\newblock Declarative vs rule-based control for flocking dynamics.
\newblock In {\em Proceedings of SAC 2018, 33rd Annual ACM Symposium on Applied
  Computing}, pages 816--823, 2018.

\bibitem{CAMACHO2007}
Eduardo~F. Camacho and Carlos Bordons~Alba.
\newblock {\em Model Predictive Control}.
\newblock Springer, 2007.

\bibitem{zhang2015model}
Hai-Tao Zhang, Zhaomeng Cheng, Guanrong Chen, and Chunguang Li.
\newblock Model predictive flocking control for second-order multi-agent
  systems with input constraints.
\newblock {\em IEEE Transactions on Circuits and Systems I: Regular Papers},
  62(6):1599--1606, 2015.

\bibitem{zhan2013flocking}
Jingyuan Zhan and Xiang Li.
\newblock Flocking of multi-agent systems via model predictive control based on
  position-only measurements.
\newblock {\em IEEE Transactions on Industrial Informatics}, 9(1):377--385,
  2013.

\bibitem{Kerrigan2000softconstraints}
Eric~C. Kerrigan and Jan~M. Maciejowski.
\newblock Soft constraints and exact penalty functions in model predictive
  control.
\newblock In {\em Proc. UKACC International Conference Control}, 2000.

\bibitem{Fletcher1987}
Roger Fletcher.
\newblock {\em Practical Methods of Optimization}.
\newblock John Wiley \& Sons, New York, NY, USA, second edition, 1987.

\bibitem{lstm}
Sepp Hochreiter and J\"{u}rgen Schmidhuber.
\newblock Long short-term memory.
\newblock {\em Neural Comput.}, 9(8):1735--1780, November 1997.

\bibitem{Everettlstm}
Michael Everett, Yu~Fan Chen, and Jonathan~P How.
\newblock Motion planning among dynamic, decision-making agents with deep
  reinforcement learning.
\newblock {\em 2018 IEEE/RSJ International Conference on Intelligent Robots and
  Systems (IROS)}, pages 3052--3059, 2018.

\bibitem{chenlstm}
Changan Chen, Yuejiang Liu, Sven Kreiss, and Alexandre Alahi.
\newblock Crowd-robot interaction: Crowd-aware robot navigation with
  attention-based deep reinforcement learning.
\newblock {\em International Conference on Robotics and Automation (ICRA)},
  2019.

\bibitem{adamopt}
Diederik~P. Kingma and Jimmy Ba.
\newblock Adam: {A} method for stochastic optimization.
\newblock In {\em 3rd International Conference on Learning Representations,
  {ICLR} 2015, San Diego, CA, USA, May 7-9, 2015, Conference Track
  Proceedings}, 2015.

\bibitem{chollet2015keras}
Fran\c{c}ois Chollet et~al.
\newblock Keras, 2015.

\bibitem{dunbar2004}
William~B. Dunbar.
\newblock {\em Distributed receding horizon control of multiagent systems}.
\newblock PhD thesis, California Institute of Technology, 2004.

\bibitem{dunbar2006}
William~B. Dunbar and Richard~M. Murray.
\newblock Distributed receding horizon control for multi-vehicle formation
  stabilization.
\newblock {\em Automatica}, 42(4):549--558, 2006.

\bibitem{Bouabdallah2007}
Samir Bouabdallah.
\newblock Design and control of quadrotors with application to autonomous
  flying.
\newblock 2007.

\bibitem{bentley2018}
Koki Shimada and Peter Bentley.
\newblock Learning how to flock: Deriving individual behaviour from collective
  behaviour with multi-agent reinforcement learning and natural evolution
  strategies.
\newblock In {\em Proceedings of the Genetic and Evolutionary Computation
  Conference Companion}, ACM, pages 169--170, 2018.

\bibitem{La2015predavoid}
Hung~Manh La, Ronny Lim, and Weihua Sheng.
\newblock Multirobot cooperative learning for predator avoidance.
\newblock {\em IEEE Transactions on Control Systems Technology}, 23(1):52--63,
  2015.

\bibitem{olfati2006flocking}
Reza Olfati-Saber.
\newblock Flocking for multi-agent dynamic systems: Algorithms and theory.
\newblock {\em IEEE Transactions on automatic control}, 51(3):401--420, 2006.

\bibitem{kahn2017RL}
Gregory Kahn, Adam Villaflor, Vitchyr Pong, Pieter Abbeel, and Sergey Levine.
\newblock Uncertainty-aware reinforcement learning for collision avoidance.
\newblock {\em arXiv preprint arXiv:1702.01182}, pages 1--12, 2017.

\bibitem{PfeifferSNSC17}
Mark Pfeiffer, Michael Schaeuble, Juan~I. Nieto, Roland Siegwart, and Cesar
  Cadena.
\newblock From perception to decision: {A} data-driven approach to end-to-end
  motion planning for autonomous ground robots.
\newblock In {\em 2017 {IEEE} International Conference on Robotics and
  Automation, {ICRA} 2017, Singapore, Singapore, May 29 - June 3, 2017}, pages
  1527--1533, 2017.

\bibitem{ros}
Morgan Quigley, Ken Conley, Brian~P. Gerkey, Josh Faust, Tully Foote, Jeremy
  Leibs, Rob Wheeler, and Andrew~Y. Ng.
\newblock Ros: an open-source robot operating system.
\newblock In {\em ICRA Workshop on Open Source Software}, 2009.

\bibitem{Godoy2016}
Julio Godoy, Ioannis Karamouzas, Stephen~J. Guy, and Maria Gini.
\newblock Moving in a crowd: Safe and efficient navigation among heterogeneous
  agents.
\newblock In {\em Proceedings of the Twenty-Fifth International Joint
  Conference on Artificial Intelligence}, IJCAI'16, pages 294--300. AAAI Press,
  2016.

\bibitem{Helwa2017}
Mohamed~K. Helwa and Angela~P. Schoellig.
\newblock Multi-robot transfer learning: A dynamical system perspective,.
\newblock In {\em Proceedings of the IEEE Intl. Conf. on Intelligent Robots and
  Systems (IROS)}, pages 4702--4708, 2017.

\bibitem{Zhou2019}
Siqi Zhou, Mohamed~K. Helwa, Angela~P. Schoellig, Andriy Sarabakha, and Erdal
  Kayacan.
\newblock Knowledge transfer between robots with similar dynamics for
  high-accuracy impromptu trajectory tracking.
\newblock In {\em Proceedings of the 18th European Control Conference}, pages
  1--8, 2019.

\bibitem{ZhangKLA2016berkeley}
Tianhao Zhang, Gregory Kahn, Sergey Levine, and Pieter Abbeel.
\newblock Learning deep control policies for autonomous aerial vehicles with
  {MPC}-guided policy search.
\newblock In {\em 2016 {IEEE} International Conference on Robotics and
  Automation, {ICRA} 2016, Stockholm, Sweden, May 16-21, 2016}, pages 528--535,
  2016.

\end{thebibliography}

\end{document}